\documentclass[%
 reprint,
 superscriptaddress,
 amsmath,amssymb,
 aps,
 prr,
nobalancelastpage
]{revtex4-2}

\usepackage{mathtools}
\usepackage{graphicx}% Include figure files
\usepackage{dcolumn}% Align table columns on decimal point
\usepackage{bm}% bold math
\usepackage{hyperref}% add hypertext capabilities
\hypersetup{colorlinks=true,linkcolor=blue,citecolor=blue,urlcolor=blue}
\usepackage{physics}
\usepackage[T1]{fontenc}
\usepackage[utf8]{inputenc}
\usepackage{soul} % \st{Hellow world} to strike out text

\usepackage{xcolor}

\newcommand{\Vs}{V_\mathrm{shift}}
\newcommand{\vpsi}{\boldsymbol{\psi}} % au choix
\newcommand{\vphi}{\boldsymbol{\phi}} % au choix
\newcommand{\vecu}{\boldsymbol{u}} % au choix
\newcommand{\vecud}{\boldsymbol{u}^\star} % au choix
\newcommand{\Hdkmax}{\hat{\mathcal{H}}^\star_{\vec{k}_\mathrm{max}}}

\newcommand{\fakefigure}[1]% #1 = label name
{\refstepcounter{figure}\label{#1}}
\newcommand{\faketable}[1]% #1 = label name
{\refstepcounter{table}\label{#1}}
\newcommand{\fakesection}[1]% #1 = label name
{\refstepcounter{section}\label{#1}}

\begin{document}

\preprint{APS/123-QED}

\title{Low- and high-energy localization landscapes for tight-binding Hamiltonians in two-dimensional lattices}% Force line breaks with \\

\author{Luis A. \surname{Razo-L\'opez}}%
% ORCID number: 0000-0002-1308-8377
\affiliation{%
    Université Côte d’Azur, CNRS, Institut de Physique de Nice (INPHYNI), France
}%
\author{Geoffroy J. \surname{Aubry}}%
% ORCID number: 0000-0002-5399-0015
\affiliation{%
    Université Côte d’Azur, CNRS, Institut de Physique de Nice (INPHYNI), France
}%
\author{Marcel \surname{Filoche}}
% ORCID number: 0000-0001-8637-3016
\affiliation{Institut Langevin, ESPCI, PSL University, CNRS, Paris, France}
\affiliation{%
    Laboratoire  de  Physique  de  la  Matière  Condenséee, CNRS, École  Polytechnique, Institut Polytechnique de Paris, 91120 Palaiseau, France
}

\author{Fabrice \surname{Mortessagne}}%
% ORCID number: 0000-0002-4457-057X
\email{fabrice.mortessagne@univ-cotedazur.fr}
\affiliation{%
    Université Côte d’Azur, CNRS, Institut de Physique de Nice (INPHYNI), France
}%

\date{\today}% It is always \today, today,

\begin{abstract}% no more than 600 characters, including spaces
Localization of electronic wave functions in modern two-dimensional (2D) materials such as graphene can impact drastically their transport and magnetic properties. The recent localization landscape (LL) theory has brought many tools and theoretical results to understand such localization phenomena in the continuous setting, but with very few extensions so far to the discrete realm or to tight-binding Hamiltonians. In this paper, we show how this approach can be extended to almost all known 2D~lattices, and propose a systematic way of designing LL even for higher dimensions. We demonstrate in detail how this LL theory works and predicts accurately not only the locations, but also the energies of localized eigenfunctions in the low and high energy regimes for the honeycomb and hexagonal lattices, making it a highly promising tool for investigating the role of disorder in these materials.
\end{abstract}

\maketitle

\section{Introduction}

The concept of Anderson localization, initially introduced in tight-binding models in the context of condensed matter physics~\cite{Anderson1958}, has been applied since in the continuous setting to all types of waves, being quantum~\cite{Jendrzejewski2012}, classical~\cite{Hu2008, Haberko2020, Aubry2020, Yamilov2023, Scheffold2022} or even gravitational~\cite{Rothstein2013}. In this setting, the recent theory of the \emph{localization landscape} (LL)~\cite{Filoche2012} has brought new insights and methods to address the wave localization properties in systems such as gases of ultracold atoms~\cite{Pelletier2022}, disordered semiconductor alloys~\cite{Filoche2017}, or enzymes~\cite{Chalopin2019}, and have been successfully extended to Dirac fermions \citep{Lemut2020} and non scalar field theory~\citep{Herviou2020}. In this work we show that the whole machinery of the LL can be generalized to tight-binding systems for most known 1D and 2D lattices, allowing us to broaden the range of predictive scope of this fruitful approach to a large class of new 2D materials~\cite{Novoselov2016} (see also \cite{2DReview} and the related collection of papers).

\section{Tight-binding and localization landscapes}

Tight-binding models are commonly used to study perfect~\cite{Bellec2013} as well as disordered lattices~\cite{Karamlou2022, Mao2020, Liu2020}. The tight-binding Hamiltonian $\hat{\mathcal{H}}$ with on-site disorder and nearest-neighbor coupling is defined as
\begin{align}    \label{eq:tightBinding}
    (\hat{\mathcal{H}} \vpsi)_n = - t \sum_{m\in \left<n\right>} (\psi_m - \psi_n) + (V_n-b_n\,t) \,\psi_n \, 
\end{align}
where $\vpsi \equiv \left(\psi_n\right)_{n\in\left[\![1,N\right]\!]}$ is the wave function defined on the sites of the lattice (numbered from 1 to $N$ here),  $V_n$ is the on-site potential at site~$n$, $t$ is the coupling constant between neighboring sites (assumed to be constant here), $\left<n\right>$ indicates the ensemble of nearest neighbors of site~$n$, and $b_n$ is its cardinal. In the following, one will assume that $t$ has value 1, thus setting the energy unit, and that $V_n = W \nu_n$ where $\nu_n$ is an i.i.d. random variable with uniform law in $[-0.5,0.5]$, $W$ being therefore the disorder strength for a given lattice.

\begin{figure*}[t]
    \includegraphics[width=\textwidth]{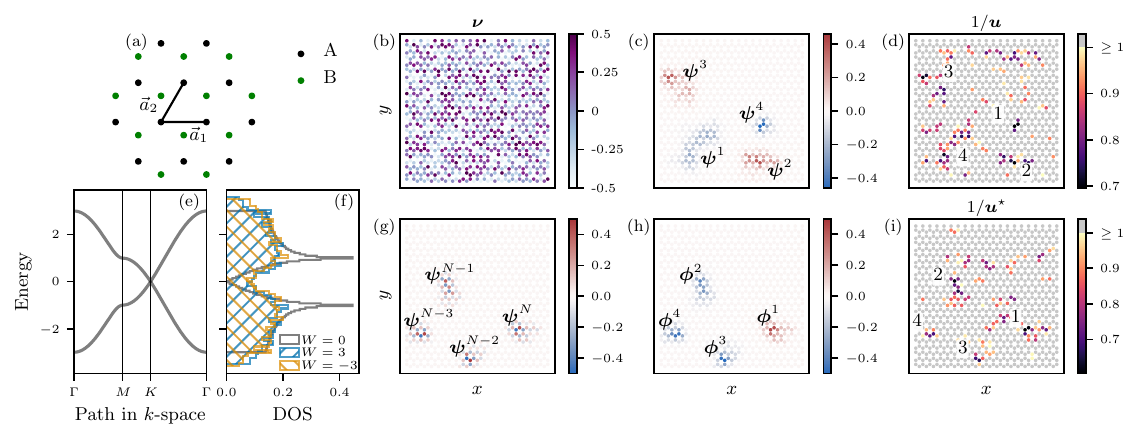}
    \caption{
    (a) The honeycomb lattice;
    (b) Plot of the random potential $V_n/W=\nu_n$;
    (c) Eigenmodes with the four lowest eigenvalues of a honeycomb lattices with on-site disorder, $N=964$ sites and $W=3$;
    (d) Inverse of the localization landscape calculated for the system as in (c) where the four lowest minima are numbered;
    (e) Band structure of the honeycomb lattice;
    (f) density of state of the honeycomb lattice without and with disorder;
    (g) Eigenmodes with the four highest eigenvalues of Eq.~\eqref{eq:tightBinding};
    (h) Eigenmodes with the four lowest eigenvalues of the inverted Hamiltonian;
    (i) Inverse of the dual landscape.
    }
    \label{fig:honeycomb}
\end{figure*}

Let us start our study by presenting results on the honeycomb lattice.
Figures~\ref{fig:honeycomb}(a) and \ref{fig:honeycomb}(e) show this lattice and its celebrated dispersion relation in the tight-binding approximation, respectively~\footnote{The band calculations shown in this work are calculated using the \href{http://www.physics.rutgers.edu/pythtb}{PythTB Python package} by S.~Coh and D.~Vanderbilt}. We solve the Schrödinger equation for the Hamiltonian defined in Eq.~(\ref{eq:tightBinding}) on the honeycomb lattice, with the on-site potential depicted in Fig.~\ref{fig:honeycomb}(b). In Fig.~\ref{fig:honeycomb}(c) are displayed the first four eigenstates which, as expected, exhibit a finite spatial extension typical of Anderson-localized modes. At the other end of the spectrum, Fig.~\ref{fig:honeycomb}(g) illustrates a feature that has no continuous counterpart: the existence of high-energy localized modes (the last four eigenstates are displayed in the example). This phenomenon is well known for instance in the case of 3D Anderson localization on a cubic lattice at low disorder strength, in which the spectrum of the Hamiltonian is symmetric in the range $\left[-6-W/2;6 + W/2\right]$ and exhibits a transition (the \emph{mobility edge}) between localized and delocalized states at both ends~\cite{Kroha1990}.

In the following, we show how to build the two discrete localization landscapes displayed in Fig.~\ref{fig:honeycomb}(d) and \ref{fig:honeycomb}(i). These landscapes allow us to accurately predict the location of the localized modes near the two band edges (low and high energy), as well as their energies, without solving Eq.~\eqref{eq:tightBinding}.
We then generalize this method to a wide class of 2D lattices.

Let us first summarize the salient features of the LL~theory in the continuous setting. For any positive definite Hamiltonian $\mathcal{H}$ in such setting, the localization landscape $u$ is defined as the solution to
\begin{align}\label{eq:landscape}
    \mathcal{H} u = 1 \,.
\end{align}
One of the main results of the LL theory is that the quantity $1/u$---which has the dimension of an energy---acts as an effective potential confining in its wells the localized states at low energy~\cite{Arnold2016}. Moreover, the energy of the local fundamental state inside each well was found to be almost proportional to the value of the potential $1/u$ at its minimum inside the well~\cite{Arnold2019},
\begin{align}\label{eq:proportionality}
    E \approx \left(1+\frac{d}{4}\right) \, \min(1/u),
\end{align}
where $d$ is the embedding dimension of the system.
$1/u$ can be understood as a renormalization of the disorder at the self-adapted local scale.

In the tight-binding setting, \citet{Lyra2015} have studied a 1D linear chain with nearest-neighbor coupling and have shown that the positions of the localized modes are given by two different localization landscapes. The low energy LL is obtained by solving the analog of Eq.~\eqref{eq:landscape} in the discrete setting, i.e., $\hat{\mathcal{H}} \vecu = \mathbf{1}$ ($\vecu \equiv \left(u_n\right)_{n\in\left[\![1,N\right]\!]}$, $\mathbf{1}$ is a vector filled with 1) with the same boundary conditions than the eigenvalue problem. Another LL, called the \emph{dual localization landscape} (DLL), gives the position of the envelope of the highly oscillating, high-energy, wave functions. More recently, \citet{Wang2021} have proved mathematically that in any dimension, the reciprocals of these discrete LL and DLL act indeed as effective confining potentials in a tight-binding system at both low- and high-energy regimes, respectively.

Figure~\ref{fig:honeycomb}(d) shows the reciprocal of the LL, $1/\vecu \equiv \left(1/u_n\right)_{n\in\left[\![1,N\right]\!]}$ computed on the honeycomb lattice with the on-site disorder depicted in Fig.~\ref{fig:honeycomb}(b). Note that a shift $\hat{\mathcal{H}}  \to \hat{\mathcal{H}} +\Vs$ has been performed in \eqref{eq:tightBinding} to ensure a positive definite Hamiltonian, see Appendix~\ref{sec:Vshift}. As already observed for continuous systems, the role of effective confining potential played by $1/\vecu$ is revealed through its basins, labelled following their depth in Fig.~\ref{fig:honeycomb}(d). Indeed, one can observe the correspondence between the deepest wells of $1/\vecu$ and the positions of the first eigenmodes plotted in Fig.~\ref{fig:honeycomb}(c). As analyzed by \citet{Arnold2019} in the continuous setting, two almost-equal eigenvalues can lead to a different ordering in the values of the minima of $1/\vecu$, thus inducing a mismatch in the correspondence. This effect, which does not affect the ability of the LL to predict the position of localized modes, is visible in Fig.~\ref{fig:honeycomb}(c) and (d) with the first and fourth eigenstates and minima, and has been analyzed in detail in the Appendix~\ref{sec:quality}. Finally, we have quantitatively tested that, regardless of the lattice, the tight-binding LL efficiently pinpoint the localized modes (see Appendix~\ref{sec:simulationsDetails}).

\section{High-energy landscape for symmetrical DOS}

On the other hand, the high-energy situation is much more intricate.
The symmetry of the honeycomb lattice allows us a straightforward derivation of the landscape governing the high-energy localized states, namely the DLL. Indeed, the tight binding Hamiltonian~\eqref{eq:tightBinding} can be decomposed into $\hat{\mathcal{H}}=\hat{\mathcal{H}}_0+ \hat{\mathcal{V}}$, where $\hat{\mathcal{H}}_0$ stands for the uniform honeycomb lattice with zero on-site energy, and $\hat{\mathcal{V}}$ accounts for the disordered on-site potential. The unperturbed part of the Hamiltonian displays the usual chiral symmetry for a bipartite lattice, $\Sigma_z\hat{\mathcal{H}}_0\Sigma_z=-\hat{\mathcal{H}}_0$, where the Pauli-like matrix $\Sigma_z$ acts on the sublattice degree of freedom: it keeps the amplitudes on the $A$ sites fixed but inverts those on the $B$ sites ($\Sigma_z=P_A-P_B$, the difference between the respective projectors on the two sub-lattices). Due the \emph{diagonal} nature of the disordered potential, the complete Hamiltonian obeys the symmetry $\Sigma_z\left(\hat{\mathcal{H}}_0+\hat{\mathcal{V}}\right)\Sigma_z=-\left(\hat{\mathcal{H}}_0-\hat{\mathcal{V}}\right)$. The latter property is exemplified in Fig.~\ref{fig:honeycomb}(e) and (f): unlike the DOS of the uniform lattice, the DOS of a given realization of the disordered system is not symmetric with respect to the origin, but the DOS obtained by inverting the sign of all on-site energies is the exact symmetric of the original situation. 

Let us call $\vphi \equiv \left(\phi_n\right)_{n\in\left[\![1,N\right]\!]}$ the eigenstates of the inverted Hamiltonian ordered by increasing eigenvalues. The low-energy states of the inverted Hamiltonian now correspond to the high-energy states of the original Hamiltonian through $\vphi =\Sigma_z\vpsi$. Since the high-energy eigenstates oscillate with a period equal to the nearest-neighbor distance, the new low-energy states appear as ``demodulated'' versions of their high-energy counterparts, see Fig.~\ref{fig:honeycomb}(g) and (h). We can now therefore use the localization landscape for the inverted system, but with $\vecud$ being now the solution to $\hat{\mathcal{H}}^\star \vecud = \boldsymbol{1}$ with
\begin{equation}
(\hat{\mathcal{H}}^\star \vphi)_n = - \sum_{m\in \left<n\right>} (\phi_m - \phi_n) +(\Vs- V_n) \,\phi_n.
\end{equation}
In the example of Fig.~\ref{fig:honeycomb}(i), one can clearly see how the deepest wells (which are different from the low-energy wells) pinpoint the locations of the localized states. Beyond the honeycomb lattice, this spectrum inversion strategy can be deployed for others lattices with symmetric band structure, as the 1D dimer chain or the 2D Lieb lattice [see Appendix Table~\ref{tab:latticeCollection}].
 
\begin{figure}
    \centering
    \includegraphics{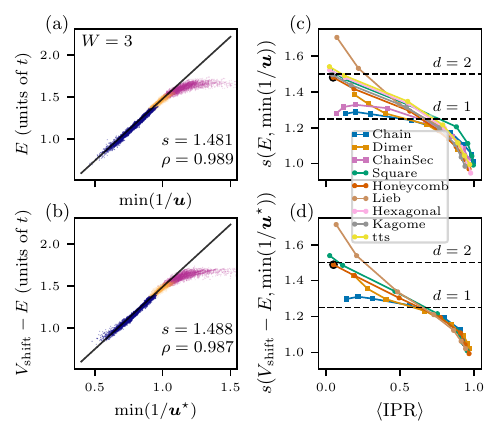}
    \caption{
    (a) Proportionality between $\mathrm{min}\left(1/\vecu\right)$ and $E$ for the lowest energies. The blue dots correspond to the 3\%~states of lowest energy for 100 different configurations, the orange dots to the 3-5\% tier, and the pink dots to the 5-7\% tier, respectively. The black line corresponds to a linear fit of the pink dots, the slope $s$ and the Pearson coefficient $\rho$ being given in the frame. (b) Proportionality between $\mathrm{min}\left(1/\vecud\right)$ and $(\Vs-E)$ for the highest energies. Similar plot to (a), but for the states of highest energy. (c)~Slope $s$ for the low-energy states, for different lattices in 1D (squares) and 2D (circles). Each symbol corresponds to a disorder strength $W$, but instead of reporting $W$ on the horizontal axis, we chose to use the average value of the IPR which is a better comparison parameter across different lattices. The dashed horizontal lines show the limits expected in the continuous case from Eq.~\eqref{eq:proportionality}. The black circle corresponds to the case displayed in (a). (d)~Similar plot to (c) for the highest-energy states.
    }
    \label{fig:proportionality}
\end{figure}

As mentioned in the introduction, the localization landscape also provides accurate estimates of the localized eigenvalues in the continuous setting~\cite{Arnold2019}. However, the generalization of the simple Eq.~\eqref{eq:proportionality} to tight-binding Hamiltonians has never been studied beyond the simple $d$-cubic lattice, nor its extension to the higher part of the spectrum. We plot on Fig.~\ref{fig:proportionality}(a) (resp. \ref{fig:proportionality}(b)) the lowest (resp. highest) eigenvalues of Eq.~(\ref{eq:tightBinding}) versus the local minimum values of the effective potential $1/\vecu$ (resp. $1/\vecud$) at the position of localized eigenstates for a honeycomb lattice with $N=2135$ sites and for a given disorder $W = 3$. Each scatter plots corresponds to 100~realizations of the disordered potential. In both cases, a direct proportionality is clearly observed for the lowest part of the plots, with Pearson coefficients of the linear regression close to~0.99. A general study of the quality of the proportionality is presented in Appendix~\ref{sec:quality}. Note also that in order to obtain this proportionality (which is more than a simple linear dependency), one has to choose~$\Vs$ so that the shifted potential has a minimum value close to zero (see Appendix~\ref{sec:Vshift}).

These observations indicate that the discrete low-energy localization landscape performs as well as its continuous analog in predicting energy and spatial distribution of localized modes without resolving an eigenvalue equation. Moreover, the high-energy DLL also exhibits the same properties. For both range of energy, LL and DLL provide a good estimate of the integrated density of states, as shown in Appendix~\ref{sec:IDOS}. All these results are not restricted to the honeycomb lattice. Figures~\ref{fig:proportionality}(c) and \ref{fig:proportionality}(d) show that the proportionality is obtained for a large variety of ``canonical'' lattices (1D: chain, dimer chain, chain with second-neighbor coupling; 2D: square, honeycomb, Lieb, hexagonal, Kagome, tts~\cite{Springer2020}) and in a wide range of strength disorder. Note that to quantify the latter unequivocally for different lattices, the parameter~$W$ is not the best suited. Indeed, for a given value of~$W$, the relative weight of the potential term in~(\ref{eq:tightBinding}) compared to the kinetic term depends on the connectivity of the discrete Laplacian $\sum_m(\psi_m-\psi_n)$. The number of edges of the graph on which this operator is relying is given by the number~$b_n$ of nearest-neighbor couplings, which itself depends on the elementary motif of each given lattice. Therefore, we use a less contingent quantity, namely the inverse participation ratio (IPR) defined for a given eigenvector $\vpsi^{(j)}=\sum_n\psi_n^{(j)}\vert n\rangle$ as $\mathrm{IPR}_j=\sum_n\abs{\psi_n^{(j)}}^4/\left(\sum_n\abs{\psi_n^{(j)}}^2\right)^2$. More precisely, in Fig.~\ref{fig:proportionality}, we consider the first (c) and last (d) 3\% of the eigenstates to compute the slopes, that are plotted versus the mean IPR corresponding to the same 3\% of the eigenstates.

With one noticeable exception for the Lieb lattice, the values of the slope $s$ appear to evolve continuously between $s=1+d/4$ and $s=1$, both for the lowest and the highest eigenvalues [see Fig.~\ref{fig:proportionality}(c) and \ref{fig:proportionality}(d)]. Moreover, all the curves bunch into two smooth master curves, one for each space dimension. In the weak disorder limit, i.e. $\langle\textrm{IPR}\rangle\to 0$, that is to say when the influence of the disordered potential is small compared to the Laplacian term, one can reasonably expect that the continuous result of  Eq.~(\ref{eq:proportionality}) still holds for both the lowest and the highest part of the spectrum. This is indeed observed: the slopes fall on the $(1+d/4)$ limit for $\langle\textrm{IPR}\rangle\to 0$. In the other limit case, when the disorder is so strong that an eigenstate is localized on a single site ($\langle\textrm{IPR}\rangle\to 1$), the LL is essentially supported locally on the same site. This means that the eigenstate and the LL are locally proportional, a feature already observed in the continuous setting. Eq.~\eqref{eq:landscape} at the only site $n$ supporting the wave function therefore becomes $Hu_n = 1 \approx E u_n$, hence $E \approx 1/u_n$ and a slope $s \simeq 1$ is expected.

The definition and the properties of the low-energy LL are valid for any lattice, in any dimension, and are not restricted to nearest-neighbor coupling. We simulated thoroughly many different ``canonical'' lattices, for which details are provided in Appendix~\ref{sec:simulationsDetails}. The construction of the high-energy DLL, however, used explicitly in our case the chiral symmetry of the honeycomb lattice, hence the central symmetry of its DOS. We will show in the last part of the letter that when this property cannot be used, there remains a general procedure which consists in ``demodulating'' the Schrödinger equation around a local maximum of the dispersion relation of the Laplacian. This procedure can be applied to any lattice, leading to a DLL that relies on the specific band structure of the given lattice.

\begin{figure}[t]
    \centering
    \scalebox{1
    }{\includegraphics{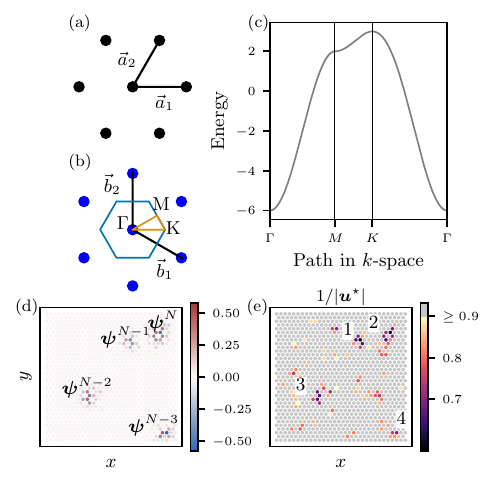}}
    \caption{
    (a) Hexagonal lattice and (b) reciprocal lattice with the first Brillouin zone and the high symmetry points.
    (c) Dispersion relation
    (d) Four states with the highest eigenvalues of a hexagonal lattice with on-site disorder, $N=1068$ and $W=3$;
    (e) Effective potential where the 4 deepest minima are marked.
    }
    \label{fig:hexagonal}
\end{figure}%

\section{High-energy landscape for non symmetrical DOS}

To illustrate this, we now focus on the hexagonal lattice [Fig.~\ref{fig:hexagonal}(a)] whose DOS is not symmetric [Fig.~\ref{fig:hexagonal}(c)]. In Fig.~\ref{fig:hexagonal}(d), we display the states corresponding to the 4 highest eigenvalues. Similarly to what was observed for the honeycomb lattice [Fig.~\ref{fig:honeycomb}(g)], we first note that the high-energy eigenstates are spatially oscillating with a period equal to the nearest-neighbor distance. Without on-site energy disorder, the highest eigenvalues are located at the vertices~$K$ of the \emph{first} Brillouin zone (BZ) [Fig.~\ref{fig:hexagonal}(b) and \ref{fig:hexagonal}(c)].
The corresponding wave vectors are $\vec{k}_\mathrm{max} = \pm \frac{4\pi}{3a}\hat{x}, \pm \left(\frac{2\pi}{3a}\hat{x} + \frac{2\pi}{\sqrt{3}a}\hat{y}\right), \pm \left(\frac{2\pi}{3a}\hat{x} - \frac{2\pi}{\sqrt{3}a}\hat{y}\right)$. To remove the rapidly oscillating part of the high-energy eigenstates, we define the envelope $\vphi$ of an eigenmode $\vpsi$ whose wave vector is close to $\vec{k}_\mathrm{max}$ by $\psi_n = e^{i\vec{k}_\mathrm{max}\cdot \vec{r}_n} \phi_n$, where $\vec{r}_n$ is the position of site $n$ (see Appendix~\ref{sec:localMaximum} for the derivation of the landscape around a local maximum of the dispersion relation).
By injecting $\vphi$ in Eq.~\eqref{eq:tightBinding}, we obtain a ``demodulated'' equation that reads for $\vec{k}_\mathrm{max} = \frac{4\pi}{3a}\hat{x}$:
\begin{equation}
    \begin{split}
        & e^{i\frac{4\pi}{3}} \phi_{n+\vec{a}_1}
        +e^{-i\frac{4\pi}{3}} \phi_{n-\vec{a}_1}
        +e^{i\frac{2\pi}{3}} \phi_{n+\vec{a}_2}\\
        +&e^{-i\frac{2\pi}{3}} \phi_{n-\vec{a}_2}
        +e^{i\frac{2\pi}{3}} \phi_{n+\vec{a}_1-\vec{a}_2}
        +e^{-i\frac{2\pi}{3}} \phi_{n-\vec{a}_1+\vec{a}_2}\\
    \MoveEqLeft[-6] -W\nu_n \phi_n = -E\phi_n,
    \label{eq:demodulationTri}
    \end{split}
\end{equation}
where, in addition, the band structure has been inverted by changing the signs of both the couplings and the on-site energies, and where the notation $n+\vec{a}$ denotes the site reached from site $n$ by a translation of a vector $\vec{a}$ [see Fig.~\ref{fig:hexagonal}(a) for the definition of $\vec{a}_1$ and $\vec{a}_2$]. Even though Eq.~\eqref{eq:demodulationTri} is a complex equation, the eigenvalues are all real as they are also the eigenvalues of the original problem~\eqref{eq:tightBinding}. The operator on the left-hand side is then made positive-definite by adding the appropriate shift~$\Vs$, leading to the definition of $\hat{\mathcal{H}}^\star_{\vec{k}_\mathrm{max}}$ of which $\vphi$ is an eigenfunction:
\begin{equation}\label{eq:DLL_asym}
\hat{\mathcal{H}}^\star_{\frac{4\pi}{3a}\hat{x}} \,\vphi= (\Vs-E) \,\vphi\,.
\end{equation}
We then compute the landscape $\vecud$ associated to this Hamiltonian, i.e. $\hat{\mathcal{H}}^\star_{\frac{4\pi}{3a}\hat{x}} \vecud= \boldsymbol{1}$, and obtain a complex confining potential $1/\vecud$ whose absolute value is plotted in Fig.~\ref{fig:hexagonal}(e). The comparison between the deepest wells of $1/\abs{\vecud}\equiv \left(1/\abs{u_n^\star}\right)_{n\in\left[\![1,N\right]\!]}$, see Fig.~\ref{fig:hexagonal}(d), and the locations of the aforementioned localized high-energy states clearly shows here again a direct match between the two sets (see Figs~\ref{fig:localizationPredictionSmallDisorder} and \ref{fig:localizationPredictionLargeDisorder} in Appendix~\ref{sec:quality}). Moreover, the proportionality between the minima of $1/\abs{\vecud}$ in the basins and the actual energies still holds (see Fig~\ref{fig:proportionalityExplicitDemodulation} in Appendix~\ref{sec:quality}). We have performed extensive simulations on 8~types of lattices (3 1D and 5 2D) for various disorder strengths, all exhibiting Pearson coefficients larger than 0.96 and even larger than 0.98 in most cases.

\section{Conclusion}

Born a decade ago, the localization landscape theory has proven its remarkable efficiency to bring in a more accessible form the information contains in a Hamiltonian~\cite{Shama00}. In this work, we have extended its scope to discrete systems described by a tight-binding Hamiltonian, with a focus on 2D lattices. The low-energy part of the spectrum is described by a discrete extension of the effective confining potential defined for continuous systems. It bears the same efficiency than its continuous counterpart in predicting the localization regions and the corresponding energies, hence the density of states~\citep{Arnold2022}. More challenging is the construction of the \emph{dual} confining potential that acts on the upper part of the spectrum. When the lattice is equipped with chiral symmetry, like the honeycomb lattice, the high-energy theory is directly deduced from the low one. When this symmetry is not present, we have proposed a general procedure to build the dual localization landscape. Our method is efficient, robust and very general but not yet completely universal. It has yet to be extended to situations like the one encountered with the Kagome lattice. In this case, the DOS is not symmetric, and the high energy states lie on a flat band: the definition of $\vec{k}_\mathrm{max}$ remains a challenge.
Finally, it is also worth noting that interesting properties appear in the center of the band for many of the new 2D materials. Although this energy range doesn't fall directly into the frame of our approach, we were able to pinpoint the localization regions in this case thanks to an approach inspired by the {${\mathcal{L}}^{2}$}-landscape method~\cite{Herviou2020}. This last one is able to give information about the localization subregions, but predictions about the eigenenergies, IDOS, etc. are still missing despite of its efficient numerical implementation~\cite{kakoi2023}. Predicting the eigenenergies and the IDOS will be the topic of future investigations.

\begin{acknowledgments}
This work was supported by a grant from the Simons Foundation (1027116, M.F.). L.A.R.-L. gratefully acknowledges the financial support from CONACyT (Mexico), through the Grant No.775585, and from the French government, through the UCA\textsuperscript{JEDI} Investments in the Future project managed by the National Research Agency (ANR) with the reference number ANR-15-IDEX-01.
\end{acknowledgments}

\appendix

\section{Energy shift}
\label{sec:Vshift}

For the simulations presented in the paper, we considered the Hamiltonian given by Eq.~(\ref{eq:tightBinding}) in the main body of the paper, but written in a more concise way:
\begin{align}
    (\hat{\mathcal{H}} \psi)_n = - t \sum_{m\in \{\mathrm{nn}\}} \psi_m + (V_n+\Vs) \psi_n \,,
\end{align}
where $V_n$ is the on-site potential at site~$n$, and $t$ is the coupling constant between neighboring sites. Additionally, $V_n = W \nu_n$ where $\nu_n$ is an i.i.d. random variable with uniform law in $[-0.5,0.5]$, $W$ being therefore the disorder strength for a given lattice, and $\Vs$ the energy shift that avoids negative eigenvalues. Finally $\Vs=-\min(\mathcal{E}) + W/2$ where $\mathcal{E}$ is the energy of the system \emph{without} disorder at \emph{zero} on-site energy. The considered values are shown in Table~\ref{tab:V0}.

To compute the dual landscape, we use $\Vs=\max(\mathcal{E})+W/2$. Note that for the 1D chain with 2\textsuperscript{nd} neighbors coupling the energy is
\begin{align}
    \mathcal{E}=-2\left[t_1\cos{\left(k_xa\right)}+t_2\cos{\left(2k_xa\right)}\right],
\end{align}
where the corresponding $k_{\mathrm{max}}$ is the solutions of
\begin{align}
    t_1\sin{\left(k_xa\right)}=-2t_2\sin{\left(2k_xa\right)}.
\end{align}
Then, values shown in Table~\ref{tab:V0} correspond to the particular case $t_2=t_1/\sqrt{8}$.

\begin{table}[h]
    \centering
    \begin{tabular}{ccc}
        Lattice & $-\min(\mathcal{E})$ & $\max(\mathcal{E})$ \\
        \hline
        1D Chain & $2$ & $2$ \\
        1D dimer chain & $t_1+t_2$ & $t_1+t_2$\\
        1D chain with 2\textsuperscript{nd} neighbors coupling & $2\left(t_1+t_2\right)$ & $\sqrt{2}t_1$\\
        Square & $4$ & $4$ \\
        Lieb & $\sqrt{8}$ & $\sqrt{8}$ \\
        tts & $5$ & $3$\\
        Hexagonal & $6$ & $3$ \\
        Honeycomb & $3$ & $3$ \\
        Kagome & $4$ & $2$
    \end{tabular}
    \caption{Smallest and largest eigenvalues of the tight binding Hamiltonian without on-site potential and with $\Vs=0$.
    The couplings are all $t=1$, except for 1D dimer chain and the 1D chain with 2\textsuperscript{nd} neighbor coupling cases where they are explicitly written.}
    \label{tab:V0}
\end{table}

\section{Accuracy of the predictions}
\label{sec:quality}

\begin{figure}
    \centering
    \includegraphics{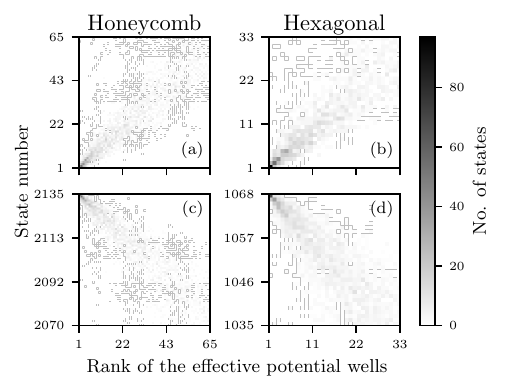}
    \caption{
    The number of states whose position match with the maxima of the landscape [(a) and (b)] and with the maxima of the dual landscape [(c) and (d)]. We consider the $3\%$ of the states for each lattice, the weakest disorder shown in Table~\ref{tab:simulations} ($W=6$ and $W=3$ for the Hexagonal and Honeycomb lattice, respectively) and 100 disorder configurations. 
    }
    \label{fig:localizationPredictionSmallDisorder}
\end{figure}

\begin{figure}
    \centering
    \includegraphics{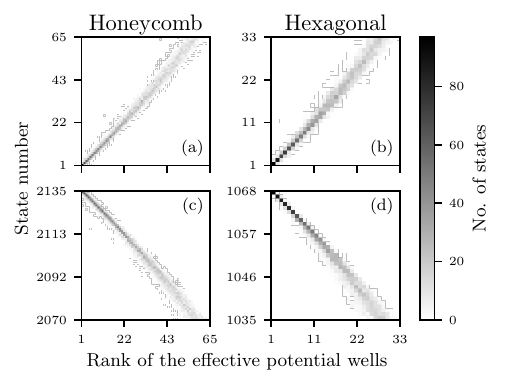}
    \caption{
    The number of states whose position match with the maxima of the landscape [(a) and (b)] and with the maxima of the dual landscape [(c) and (d)]. We consider the $3\%$ of the states for each lattice, the strongest disorder shown in Table~\ref{tab:simulations} ($W=480$ and $W=240$ for the Hexagonal and Honeycomb lattice, respectively) and 100 disorder configurations.
    }
    \label{fig:localizationPredictionLargeDisorder}
\end{figure}

In this section, we quantify the quality of the localization prediction of the localized states in a situation of small (Fig.~\ref{fig:localizationPredictionSmallDisorder}) and large disorder (Fig.~\ref{fig:localizationPredictionLargeDisorder}).
This is done by calculating the distances between the position of the maximum of an eigenstate, and the positions of all the wells of the effective potential. The wells are ordered by their depths, and we then find the rank of the well corresponding to the minimum distance, and report it in the 2D histograms of Fig.~\ref{fig:localizationPredictionSmallDisorder} and \ref{fig:localizationPredictionLargeDisorder}. In the high-energy case, the computations are done using the  symmetry of the bandstructure for the honeycomb lattices [Figs.~\ref{fig:localizationPredictionSmallDisorder}(c) and \ref{fig:localizationPredictionLargeDisorder}(c)], and using the explicit demodulation for the hexagonal lattices [Figs.~\ref{fig:localizationPredictionSmallDisorder}(d) and \ref{fig:localizationPredictionLargeDisorder}(d)].

\begin{figure}
    \centering
    \includegraphics{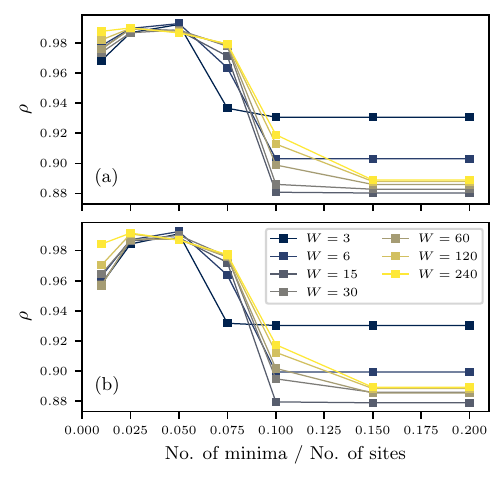}
    \caption{
    Pearson correlation coefficient as a function of number of minima taken into account for the honeycomb lattice with different strengths of disorder $W$;
    (a) The landscape prediction;
    (b) The dual landscape prediction using the symmetry of the bandstructure;
    The linear regression quality do not evolve when the number of minima considered is larger than the actual number of minima of $1/\vecu$ ($1/\vecud$).
    }
    \label{fig:PearsonHoneycomb}
\end{figure}

\begin{figure}
    \centering
    \includegraphics[width=\columnwidth]{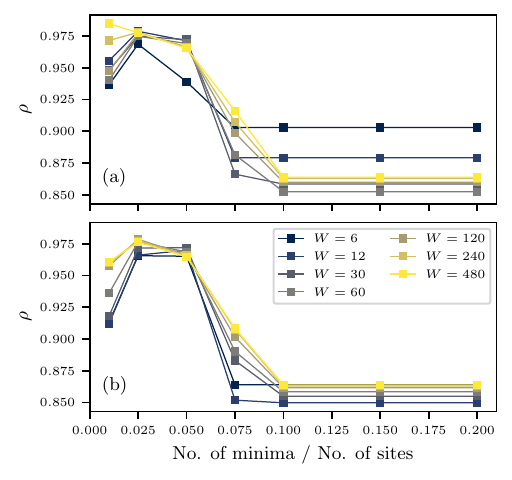}
    \caption{
    Pearson correlation coefficient as a function of number of minima taken into account for the hexagonal lattice with different strengths of disorder $W$;
    (a) The landscape prediction;
    (b) The dual landscape prediction using the explicit demodulation;
    The linear regression quality do not evolve when the number of minima considered is larger than the actual number of minima of $1/\vecu$ ($1/\abs{\vecud}$).
    }
    \label{fig:PearsonHexagonal}
\end{figure}

Simmilarly, we quantify energy predictions for all the methods presented in the Main Text.
In Fig.~\ref{fig:PearsonHoneycomb}, we plot the Pearson coefficient of the linear regression as a function of the number of minima taken to compute the slope for the honeycomb lattice, and in Fig.~\ref{fig:PearsonHexagonal} the same quantity for the hexagonal lattice.
We see that whatever the disorder, the maximum correlation is obtained when we chose the 3\% states with the lowest (or highest) eigenvalues.

\begin{figure}
    \centering
    \includegraphics[width=\columnwidth]{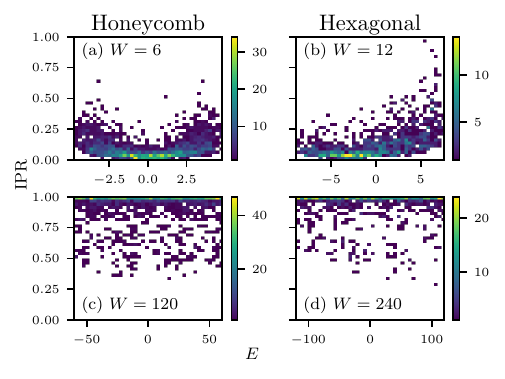}
    \caption{
    2D histogram showing how the eigenvalues and the IPR are distributed.
    }
    \label{fig:IPR_energy}
\end{figure}

In Fig.~\ref{fig:IPR_energy}, we plot a 2D histogram showing the distribution of the eigenstates in the eigenvalues-IPR plane.
We clearly see the more localized states for higher disorder, and the asymmetry of the density of states for the hexagonal lattices.

\begin{figure}
    \centering
    \includegraphics[width=\columnwidth]{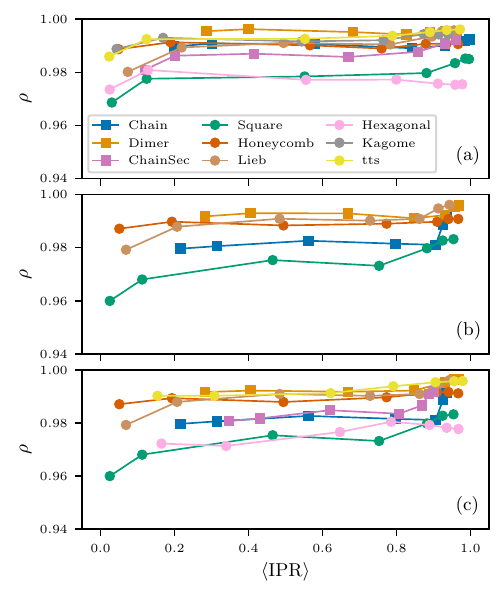}
    \caption{
    Pearson correlation coefficient as a function of the $\expval{\textrm{IPR}}$ corresponding to the data plotted in Main Text Fig.~\ref{fig:proportionality};
    (a) The landscape prediction;
    (b) The dual landscape prediction for the symmetric lattices using the symmetry of the DOS property;
    (c) The dual landscape prediction for all the lattices using the explicit demodulation.
    }
    \label{fig:IPR_Pearson}
\end{figure}

Figure~\ref{fig:IPR_Pearson} shows that in every case considered in the paper, the eigenvalues and the minima of the effective potential are highly correlated (Pearson coeeficient very close to 1).

\begin{figure}
    \centering
    \includegraphics{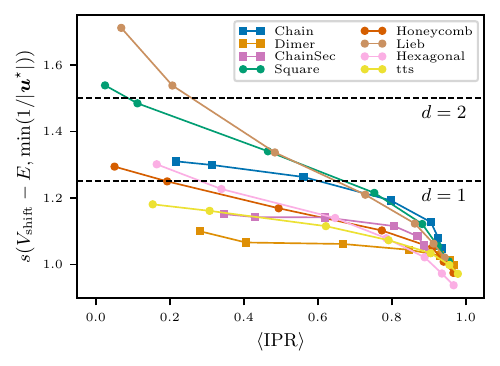}
    \caption{
    Proportionality factor between $\Vs-E$ and the minima of $1/\abs{\vecud}$ for the different 1D (squares) and 2D (circles) lattices studied. The linear fits are done on states with the 3\% highest energy. The dashed horizontal lines show the limit expected using the approximate form Eq.~\eqref{eq:proportionality}. Each symbol correspond to a disorder strength. The results are plotted as a function of the mean IPR calculated over the 3\% states found for a given disorder strength.
    }
    \label{fig:proportionalityExplicitDemodulation}
\end{figure}

Finally, Fig.~\ref{fig:proportionalityExplicitDemodulation} is similar to Main Text Fig.~\ref{fig:proportionality}(d), except that the calculations are done using the explicit demodulation.
This explains why there are more cases in Fig.~\ref{fig:proportionalityExplicitDemodulation} than in Fig.~\ref{fig:proportionality} as the constraint on the symmetry of the DOS is lifted.

\begin{table*}[h]
    \centering
    \begin{tabular}{ccccc}
        Lattice & Bravais lattice & $\sharp$ of bands & DOS Sym? & $k_\mathrm{max}a$ \\
        \hline
        1D chain & 1D & 1 & y & $\pm \pi$ \\
        1D dimer chain $(t_2=t_1/2)$ & 1D & 2 & y & $\pm 2\pi$ \\
        1D chain with 2\textsuperscript{nd} neighbour coupling & 1D & 1 & n & $\pm \frac{3\pi}{4}$ \\
        Square & sql & 1 & y & M $(\pm \pi, \pm \pi)$ \\
        Lieb & sql & 3 & y & $(\pm 2\pi, \pm2\pi)$ (flat band in the middle) \\
        tts & sql & 4 & n & $(\pm 2\pi, \pm2\pi)$ \\
        hexagonal & hxl & 1 & n & K $\left(\pm \frac{4\pi}{3},0\right)$, $\pm\left(\frac{2\pi}{3}, \frac{2\pi}{\sqrt{3}}\right)$, $\pm\left(\frac{2\pi}{3}, -\frac{2\pi}{\sqrt{3}}\right)$ \\
        Honeycomb & hxl & 2 & y & $ \left(0,\pm \frac{4\pi}{\sqrt{3}}\right)$, $\pm\left(2\pi,\frac{2\pi}{\sqrt{3}}\right)$, $\pm\left(2\pi,-\frac{2\pi}{\sqrt{3}}\right)$ \\
        Kagome & hxl & 3 & n &
        %$\left(\pm \frac{8\pi}{3},0\right)$, $\pm\left(\frac{4\pi}{3}, \frac{4\pi}{\sqrt{3}}\right)$, $\pm\left(\frac{4\pi}{3}, -\frac{4\pi}{\sqrt{3}}\right)$
        flat band
    \end{tabular}
    \caption{High energy wave vector for the different lattices studied.}
    \label{tab:latticeCollection}
\end{table*}

\begin{table*}[h]
    \centering
    \begin{tabular}{ccccc}
        Lattice & Number of sites $N$ & Disorder calculated $W$ & Number of configurations calculated \\
        \hline
        1D chain & 1001 & 1, 2, 4, 10, 20, 40, 80, 160 & 100 \\
        1D dimer chain $(t_2=t_1/2)$ & 2001 & 1, 2, 4, 10, 20, 40, 80, 160 & 100 \\
        1D chain with 2\textsuperscript{nd} neighbour coupling & 1001 & 1, 2, 4, 10, 20, 40, 80, 160 & 100 \\
        Square & 961 & 4, 8, 20, 40, 80, 160, 320 & 100 \\
        Lieb & 2821 & $\frac{8}{3}$, $\frac{16}{3}$, $\frac{40}{3}$, $\frac{80}{3}$, $\frac{160}{3}$, $\frac{320}{3}$, $\frac{640}{3}$ & 100 \\
        tts & 3661 & 5, 10, 25, 50, 100, 200, 400 & 100 \\
        Hexagonal & 1068 & 6, 12, 30, 60, 120, 240, 480 & 100 \\
        Honeycomb & 2135 & 3, 6, 15, 30, 60, 120, 240 & 100 \\
        Kagome & 3185 & 4, 8, 20, 40, 80, 160, 320 & 100 
    \end{tabular}
    \caption{Summary of the simulations}
    \label{tab:simulations}
\end{table*}

\section{Exhaustive study of various 1D and 2D~lattices and simulations details}
\label{sec:simulationsDetails}

In Table.~\ref{tab:latticeCollection}, we list the different lattices studied in this study with their properties. The values of $k_{\mathrm{max}}$ shown here for the 1D chain with 2\textsuperscript{nd} neighbor coupling correspond to the case $t_2=t_1/\sqrt{8}$. For each lattice, we computed 100 different configurations, see Table~\ref{tab:simulations} for details. In the case of symmetric DOS, we calculated the landscape prediction both using the symmetry of the DOS and using the explicit demodulation. In Fig.~\ref{fig:proportionality}(c) and (d) of the Main text, 3\% of the eigenstates correspond to 64 states, and because we consider 100 configurations, the slopes are calculated using $6\,400$ points.

\section{IDOS estimate}
\label{sec:IDOS}

\begin{figure}
    \centering
    \includegraphics{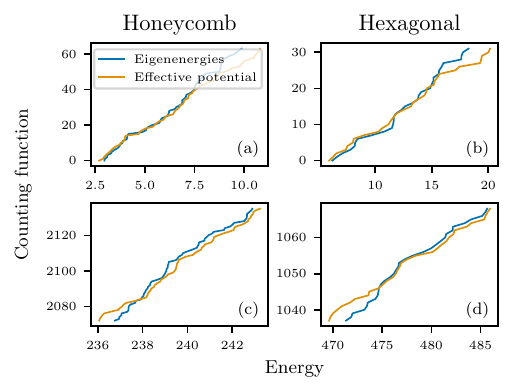}
    \caption{The counting function computed by [(a) and  (b)] the confining potential, and by [(c) and (d)] the dual confining potential. We consider the 3\% of the states for each lattice. Here, the strongest disorder was chosen (W=240 and W=480 for the Honeycomb an Hexagonal lattice, respectively).
    }
    \label{fig:IDOS}
\end{figure}

Figure \ref{fig:IDOS} shows the counting function for the Hexagonal and Honeycomb lattices with a strong on-site disorder (W=240 and W=480 for the Honeycomb an Hexagonal lattice, respectively). Blue lines correspond to the solution of Eq.~(\ref{eq:tightBinding}) while orange lines are the minima of both confining and dual confining potentials.

\section{The localization landscape around a local maximum of the dispersion relation}
\label{sec:localMaximum}

Let us assume that $\vec{k}_{\max}$ is a wave vector at which the dispersion relation $E(\vec{k})$ of the Hamiltonian without potential (i.e., minus the Laplacian) exhibits a local maximum. Then one can write locally the dispersion relation as
\begin{align}
E(\vec{k}) = ~&E(\vec{k}_{\max}) - \frac{1}{2} (\vec{k}_{\max} - \vec{k})^\top \mathcal{A} \, (\vec{k}_{\max} - \vec{k}) \nonumber \\ &+ \mathrm{o}\left(\norm{\vec{k}-\vec{k}_{\max}}^2\right) \,,
\end{align}
where $\mathcal{A}$ is a definite positive 2-by-2 matrix whose eigenvalues are the inverse of the effective masses in both directions.

One can then write any eigenfunction $\vpsi$ of the full Hamiltonian (i.e., with potential $V=W \nu$) as
\begin{equation}
\vpsi = \exp(j \vec{k}_{\max} \cdot \vec{r} ) \, \vphi
\end{equation}
where $\vphi$ is an envelope function satisfying the following equation
\begin{align}
- \sum_{m\in \{\mathrm{nn}\}} (e^{i \vec{k}_{\max} \cdot (\vec{r}_m -\vec{r}_n) } \phi_m - \phi_n) + W \nu_n \phi_n = E \phi_n \,,
\end{align}
$E$ being the energy of $\vpsi$. The local maximum of the dispersion relation $E(\vec{k})$ at $\vec{k}_{\max}$ implies that
\begin{equation}
\sum_{m\in \{\mathrm{nn}\}} \vec{r}_m \, e^{i \vec{k}_{\max} \cdot \vec{r}_m} = \vec{0} \,,
\end{equation}
where the sum is taken over all interacting neighbors of one site of the lattice assumed to be located at $\vec{r} = \vec{0}$. One can thus define a new Hamiltonian $\Hdkmax$ by:
\begin{align}
(\Hdkmax \vphi)_n = ~ & E(\vec{k}_{\max}) + \sum_{m\in \{\mathrm{nn}\}} (e^{i \vec{k}_{\max} \cdot (\vec{r}_m -\vec{r}_n) } \phi_m - \phi_n) \nonumber \\
&- W \nu_n \, \phi_n \,.
\end{align}
The energy of a plane wave $\vphi = \exp(j \vec{k} \cdot \vec{r})$ for this Hamiltonian $\Hdkmax$ is therefore
\begin{align}
E^*(\vec{k}) &= \ev{\Hdkmax}{\vphi} \nonumber \\
&= \frac{1}{2} \vec{k}^\top \mathcal{A} \, \vec{k} - \frac{W}{2} + \mathrm{o}\left(\norm{\vec{k}}^2\right) .
\end{align}
One can therefore apply the localization landscape formalism to this Hamiltonian shifted by a quantity $W/2$.

\clearpage
\bibliography{biblio}% Produces the bibliography via BibTeX.
\end{document}